\documentclass[12pt,a4paper]{article}
\pdfoutput=1
\usepackage{ifthen}
\usepackage{amssymb}
\usepackage{amsfonts}
\usepackage{upgreek}
\usepackage{xspace}
\usepackage{color}
\usepackage{amsmath}
\usepackage{graphicx}
\usepackage{bm}
\usepackage{ctable}
\usepackage{placeins}
\usepackage{grffile}
\usepackage[font=normalsize,subrefformat=parens,labelformat=parens]{subfig}
\usepackage{mathrsfs}
\usepackage{multirow}
\usepackage{relsize}
\usepackage{afterpage}
\usepackage{lmodern}
\usepackage{mciteplus}
\usepackage[bookmarks=true,colorlinks=true,linkcolor=black,citecolor=black,
urlcolor=black]{hyperref}
\usepackage[all]{hypcap}

\newcommand{\fig}[1]  {\mbox{Fig.\ #1\xspace}}

\newcommand{\equ}[1]  {\mbox{Eq.\ #1\xspace}}

\newcommand{\rfr}[1]  {\mbox{Ref.\ #1\xspace}}

\newcommand{\rfrs}[1] {\mbox{Refs.\ #1\xspace}}

\newcommand{\tab}[1]  {\mbox{Table\ #1\xspace}}

\newcommand{\lep}{\ensuremath{\ell}\xspace}
\newcommand{\had}{\ensuremath{h}\xspace}
\newcommand{\z}{\ensuremath{Z}\xspace}
\newcommand{\w}{\ensuremath{W}\xspace}
\newcommand{\ww}{\ensuremath{{WW}}\xspace}
\newcommand{\ttbar}{\ensuremath{{t\bar{t}}}\xspace}

\newcommand{\dimu}{\ensuremath{{\mu\mu}}\xspace}
\newcommand{\dilep}{\ensuremath{{\lep\lep}}\xspace}
\newcommand{\ditau}{\ensuremath{{\tau\tau}}\xspace}
\newcommand{\mumu}{\ensuremath{{\tau_\mu\tau_\mu}}\xspace}
\newcommand{\mue}{\ensuremath{{\tau_\mu\tau_e}}\xspace}
\newcommand{\emu}{\ensuremath{{\tau_e\tau_\mu}}\xspace}
\newcommand{\muh}{\ensuremath{{\tau_\mu\tau_\had}}\xspace}
\newcommand{\eh}{\ensuremath{{\tau_e\tau_\had}}\xspace}

\newcommand{\hz}{\ensuremath{{\Phi^0}}\xspace}
\newcommand{\hH}{\ensuremath{H}\xspace}
\newcommand{\hhz}{\ensuremath{{h^0}}\xspace}
\newcommand{\hAz}{\ensuremath{{A^0}}\xspace}

\newcommand{\cp}{\ensuremath{C\!P}\xspace}
\newcommand{\mhmax}{\ensuremath{{m_\hhz^\mathrm{max}}}\xspace}

\newcommand{\pb}{\ensuremath{\mathrm{pb}}\xspace}
\newcommand{\ipb}{\ensuremath{\mathrm{pb}^{-1}}\xspace}

\newcommand{\ifb}{\ensuremath{\mathrm{fb}^{-1}}\xspace}
\newcommand{\tev}{\ensuremath{\mathrm{Te\kern -0.1em V}}\xspace}
\newcommand{\tevc}{\ensuremath{\mathrm{Te\kern -0.1em V\!/}c}\xspace}
\newcommand{\tevcc}{\ensuremath{\mathrm{Te\kern -0.1em V\!/}c^2}\xspace}
\newcommand{\gev}{\ensuremath{\mathrm{Ge\kern -0.1em V}}\xspace}
\newcommand{\gevc}{\ensuremath{\mathrm{Ge\kern -0.1em V\!/}c}\xspace}
\newcommand{\gevcc}{\ensuremath{\mathrm{Ge\kern -0.1em V\!/}c^2}\xspace}
\newcommand{\mev}{\ensuremath{\mathrm{Me\kern -0.1em V}}\xspace}
\newcommand{\mevc}{\ensuremath{\mathrm{Me\kern -0.1em V\!/}c}\xspace}
\newcommand{\mevcc}{\ensuremath{\mathrm{Me\kern -0.1em V\!/}c^2}\xspace}

\newcommand{\lhcb}{LHCb\xspace}
\newcommand{\atlas}{ATLAS\xspace}
\newcommand{\cms}{CMS\xspace}
\newcommand{\feynhiggs}{{\sc FeynHiggs}\xspace}
\newcommand{\pythia}{{\sc Pythia}\xspace}

\newcommand{\dynnlo}{{\sc Dynnlo}\xspace}

\newcommand{\cteq}{{CTEQ$6$L$1$}\xspace}

\newcommand{\geant}{{\sc Geant4}\xspace}

\newcommand{\pt}{\ensuremath{p_\mathrm{T}}\xspace}

\newcommand{\eff}[2]{\ensuremath{{\varepsilon_{#1}^{#2}}}\xspace}

\newcommand{\ewk}{\ensuremath{\mathrm{EWK}}\xspace}
\newcommand{\qcd}{\ensuremath{\mathrm{QCD}}\xspace}
\newcommand{\simu}{\ensuremath{\mathrm{sim}}\xspace}

\newcommand{\cls}{\ensuremath{{\mathrm{CL_s}}}\xspace}

\newcommand{\po}{\ensuremath{\phantom{1}}\xspace}

{\end{equation}\end{linenomath*}}%

\graphicspath{{./}{Figures/}}

\begin{document}

\newcommand{\thetitle}{Limits on neutral Higgs boson production in the
  forward region in $pp$ collisions at $\sqrt{s} = {7~\mathrm{TeV}}$}

\begin{titlepage}
  
  \pagenumbering{roman}
  \vspace*{-1.5cm}
  \centerline{\large EUROPEAN ORGANIZATION FOR NUCLEAR RESEARCH (CERN)}
  \vspace*{1.5cm}
  \hspace*{-0.5cm}
  \begin{tabular*}{\linewidth}{lc@{\extracolsep{\fill}}r}
    \vspace*{-2.7cm}
    \mbox{\!\!\!\includegraphics[width=.14\textwidth]{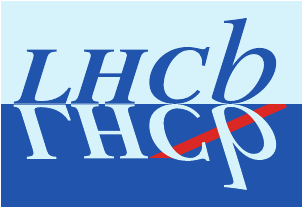}}
    & & \\
    & & CERN-PH-EP-2013-058 \\
    & & LHCb-PAPER-2013-009 \\
    & & May $27$, $2013$ \\
    & & \\
  \end{tabular*}
  
  \vspace*{3.0cm}
  {\bf\huge\boldmath
    \begin{center}
      \thetitle
    \end{center}
  }
  
  \vspace*{0.5cm}
  \begin{center}
    The \lhcb collaboration\footnote{Authors are listed on the
      following pages.}
  \end{center}
  
  \vspace*{0.5cm}
  \begin{abstract}
    \noindent Limits on the cross-section times branching fraction for neutral Higgs
bosons, produced in $pp$ collisions at ${\sqrt{s} = 7~\tev}$, and
decaying to two tau leptons with pseudorapidities between $2.0$ and
$4.5$, are presented. The result is based on a dataset, corresponding
to an integrated luminosity of $1.0~\ifb$, collected with the \lhcb
detector. Candidates are identified by reconstructing final states
with two muons, a muon and an electron, a muon and a hadron, or an
electron and a hadron. A model independent upper limit at the ${95\%}$
confidence level is set on a neutral Higgs boson cross-section times
branching fraction. It varies from ${8.6~\pb}$ for a Higgs boson mass
of ${90~\gev}$ to ${0.7~\pb}$ for a Higgs boson mass of ${250~\gev}$,
and is compared to the Standard Model expectation. An upper limit on
${\tan\beta}$ in the Minimal Supersymmetric Model is set in the \mhmax
scenario. It ranges from $34$ for a \cp-odd Higgs boson mass of
${90~\gev}$ to $70$ for a pseudo-scalar Higgs boson mass of
${140~\gev}$.

  \end{abstract}
  
  \vspace*{0.5cm}
  \begin{center}
    Published in JHEP Vol. 2013, Number 5 (2013), 132, DOI
    \href{http://dx.doi.org/10.1007/JHEP05(2013)132}{10.1007/JHEP05(2013)132}
  \end{center}
  \vspace{\fill}

  {\footnotesize \centerline{\copyright~CERN on behalf of the \lhcb
      collaboration, license
      \href{http://creativecommons.org/licenses/by/3.0/}{CC-BY-3.0}.}}
  \vspace*{2mm}

\end{titlepage}

\newpage
\setcounter{page}{2}
\phantom{}
\newpage

\centerline{\large\bf LHCb collaboration}
\begin{flushleft}
\small
R.~Aaij$^{40}$, 
C.~Abellan~Beteta$^{35,n}$, 
B.~Adeva$^{36}$, 
M.~Adinolfi$^{45}$, 
C.~Adrover$^{6}$, 
A.~Affolder$^{51}$, 
Z.~Ajaltouni$^{5}$, 
J.~Albrecht$^{9}$, 
F.~Alessio$^{37}$, 
M.~Alexander$^{50}$
% ...

{\footnotesize \it
% ...
$ ^{5}$Clermont Universit\'{e}, Universit\'{e} Blaise Pascal, CNRS/IN2P3, LPC, Clermont-Ferrand, France\\
$ ^{6}$CPPM, Aix-Marseille Universit\'{e}, CNRS/IN2P3, Marseille, France\\
% ...
$ ^{9}$Fakult\"{a}t Physik, Technische Universit\"{a}t Dortmund, Dortmund, Germany\\
% ...
$ ^{35}$Universitat de Barcelona, Barcelona, Spain\\
$ ^{36}$Universidad de Santiago de Compostela, Santiago de Compostela, Spain\\
$ ^{37}$European Organization for Nuclear Research (CERN), Geneva, Switzerland\\
% ...
$ ^{40}$Nikhef National Institute for Subatomic Physics, Amsterdam, The Netherlands\\
% ...
$ ^{45}$H.H. Wills Physics Laboratory, University of Bristol, Bristol, United Kingdom\\
% ...
$ ^{50}$School of Physics and Astronomy, University of Glasgow, Glasgow, United Kingdom\\
$ ^{51}$Oliver Lodge Laboratory, University of Liverpool, Liverpool, United Kingdom\\
% ...
$ ^{n}$LIFAELS, La Salle, Universitat Ramon Llull, Barcelona, Spain\\
% ...
$ ^{s}$Scuola Normale Superiore, Pisa, Italy\\
}
\end{flushleft}

\cleardoublepage

\setcounter{page}{1}
\pagenumbering{arabic}
\section{Introduction}

The discovery of a boson with a mass of about ${125~\gev}$ by the
\atlas~\cite{AtlasHiggs} and \cms~\cite{CmsHiggs} collaborations
requires further investigations to confirm whether its properties are
compatible with a Standard Model (SM) Higgs boson or if it is better
described by theories beyond the SM, such as supersymmetry. The ATLAS
and CMS measurements have been made at central values of
pseudorapidity, $\eta$; investigations in the forward region can be
provided by the LHCb experiment, which is fully instrumented between
${2 < \eta < 5}$.  Both measurements of cross-sections and branching
fractions allow different models to be tested. In this paper,
model-independent limits on the Higgs boson\footnote{The symbol \hz is
  used throughout to indicate any neutral Higgs boson. Additionally,
  charge conjugation is implied and the speed of light is taken as
  $1$.} cross-section times branching fraction into two tau leptons
are presented for the forward region and compared to SM Higgs boson
predictions. Model-dependent limits for the Minimal Supersymmetric
Model (MSSM) Higgs bosons, in the scenario where the lightest
supersymmetric Higgs boson mass is maximal (\mhmax)~\cite{Benchmarks},
are also given for the ratio between up- and down-type Higgs vacuum
expectation values (${\tan\beta}$) as a function of the \cp-odd Higgs
boson (\hAz) mass.

\section{Detector and datasets}

The \lhcb detector~\cite{Detector} is a single-arm forward
spectrometer. The components of particular relevance for this analysis
are a high-precision tracking system consisting of a silicon-strip
vertex detector surrounding the $pp$ interaction region, a large-area
silicon-strip detector located upstream of a dipole magnet with a
bending power of about ${4~\mathrm{Tm}}$, and three stations of
silicon-strip detectors and straw drift tubes placed downstream of the
magnet. Photon, electron and hadron candidates are identified by a
calorimeter system consisting of scintillating-pad and pre-shower
detectors, an electromagnetic calorimeter and a hadronic
calorimeter. Muons are identified by a system composed of alternating
layers of iron and multiwire proportional chambers. The
trigger~\cite{Trigger} consists of a hardware stage, based on
information from the calorimeter and muon systems, followed by a
software stage, which applies a full event reconstruction.

Simulated data samples are used to calculate signal and background
contributions, determine efficiencies, and estimate systematic
uncertainties. Each sample was generated as described in
\rfr{\cite{Tune}}, with \pythia $6.4$~\cite{Pythia} using the \cteq
leading-order PDF set~\cite{cteq} and passed through a
\geant~\cite{GeantA, *GeantB} based simulation of the
detector~\cite{Gauss}. The \lhcb reconstruction software~\cite{Brunel}
was used to perform trigger emulation and full event reconstruction.

The dataset used for this analysis is identical to that described in
our previous measurement of the \z cross-section using tau final
states~\cite{LhcbZtautau}, which corresponded to an integrated
luminosity of ${1028 \pm 36~\ipb}$, taken at a centre-of-mass energy of
${7~\tev}$. The ${\z \to \ditau}$ decays are identified in five
categories: \mumu, \mue, \emu, \muh and \eh, defined so as to be
exclusive, where the subscripts indicate tau decays containing a muon
($\mu$), electron ($e$), or hadron ($h$) and the ordering specifies
the first and second tau decay product on which different requirements
are applied. The first tau decay product is required to have
transverse momentum, \pt, above ${20~\gev}$ and the second to have
${\pt > 5~\gev}$. Both tracks are required to have pseudorapidities
between $2.0$ and $4.5$, to be isolated with little surrounding
activity, to be approximately back-to-back in the azimuthal
coordinate, and their combined invariant mass must be greater than
${20~\gev}$.  The tracks in the \mumu, \muh, and \eh categories are
required to be displaced from the primary vertex. Additionally, the
\mumu category requires a difference between the \pt of the two tracks
and excludes di-muon invariant masses between $80$ and ${100~\gev}$, to
suppress the direct decays of \z bosons into two muons. Full details
on the selection criteria can be found in \rfr{\cite{LhcbZtautau}}.

\begin{figure}
  \newlength{\plotwidth}
  \newlength{\plotsep}
  \newlength{\capwidth}
  \newlength{\capsep}
  \setlength{\plotwidth}{0.5\textwidth}
  \setlength{\plotsep}{-0.3cm}
  \setlength{\capwidth}{2.4cm}
  \setlength{\capsep}{-1.6cm}
  \begin{center}
    \vspace{-1.9cm}
    \centerline{
      \begin{tabular}{@{\extracolsep{-0.6cm}}cc}
        \subfloat[]{\label{fig:mumu}}\hspace{\capwidth} &
        \subfloat[]{\label{fig:mue}}\hspace{\capwidth} \\[\capsep]
        \includegraphics[width=\plotwidth]
        {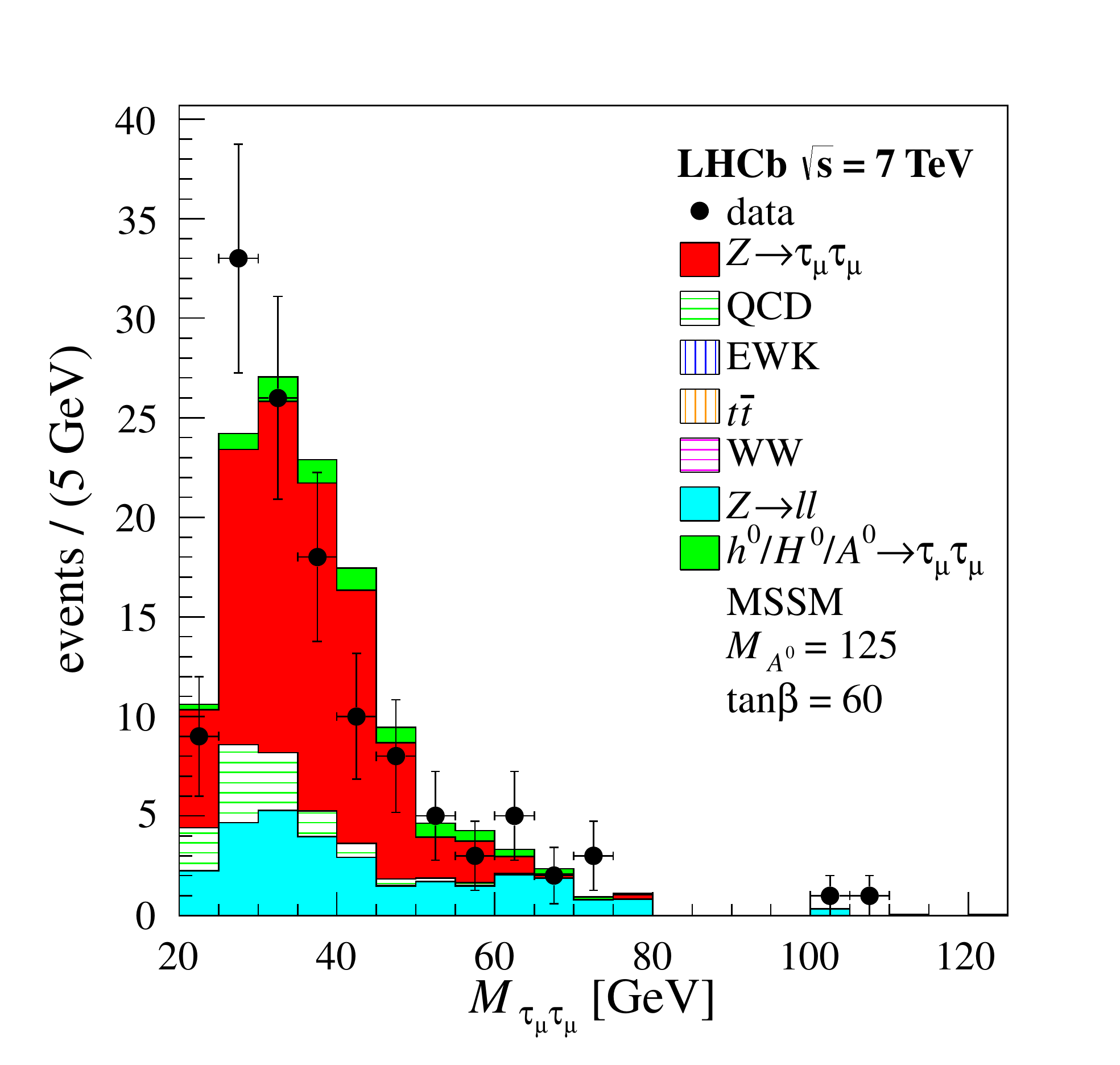} &
        \includegraphics[width=\plotwidth]
        {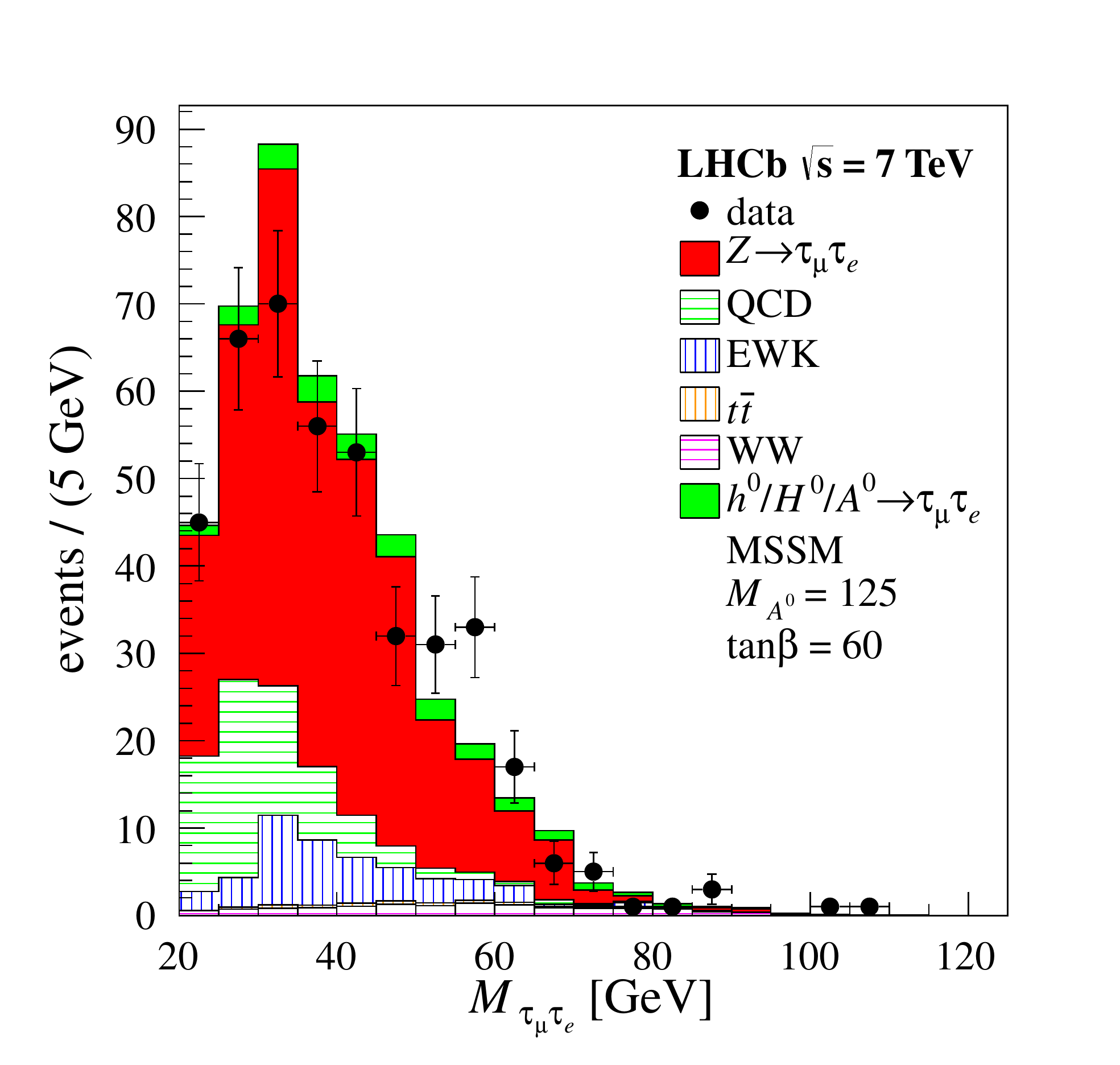}
        \\[\plotsep]
        \subfloat[]{\label{fig:emu}}\hspace{\capwidth} &
        \subfloat[]{\label{fig:muh}}\hspace{\capwidth} \\[\capsep]
        \includegraphics[width=\plotwidth]
        {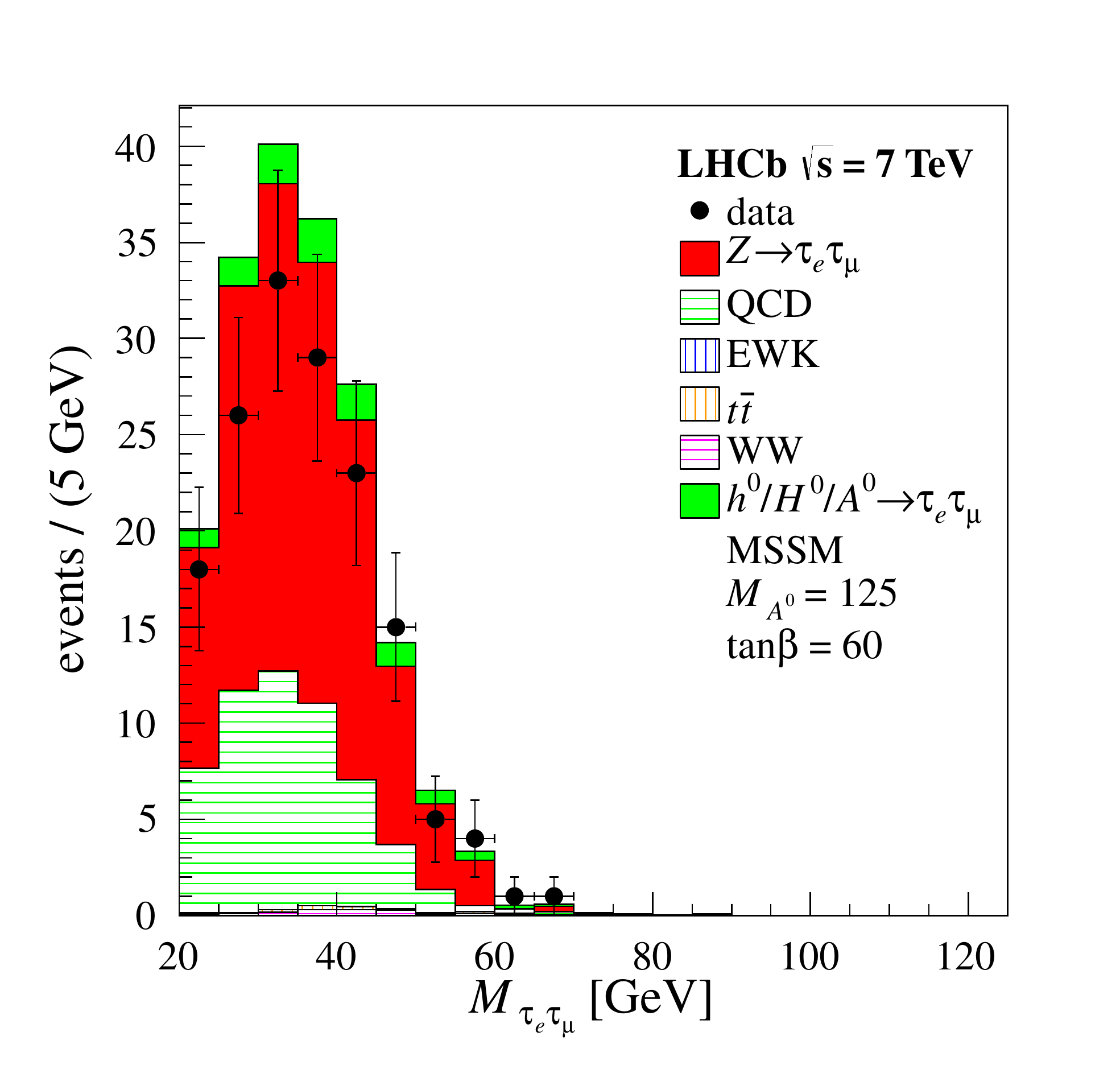} & 
        \includegraphics[width=\plotwidth]
        {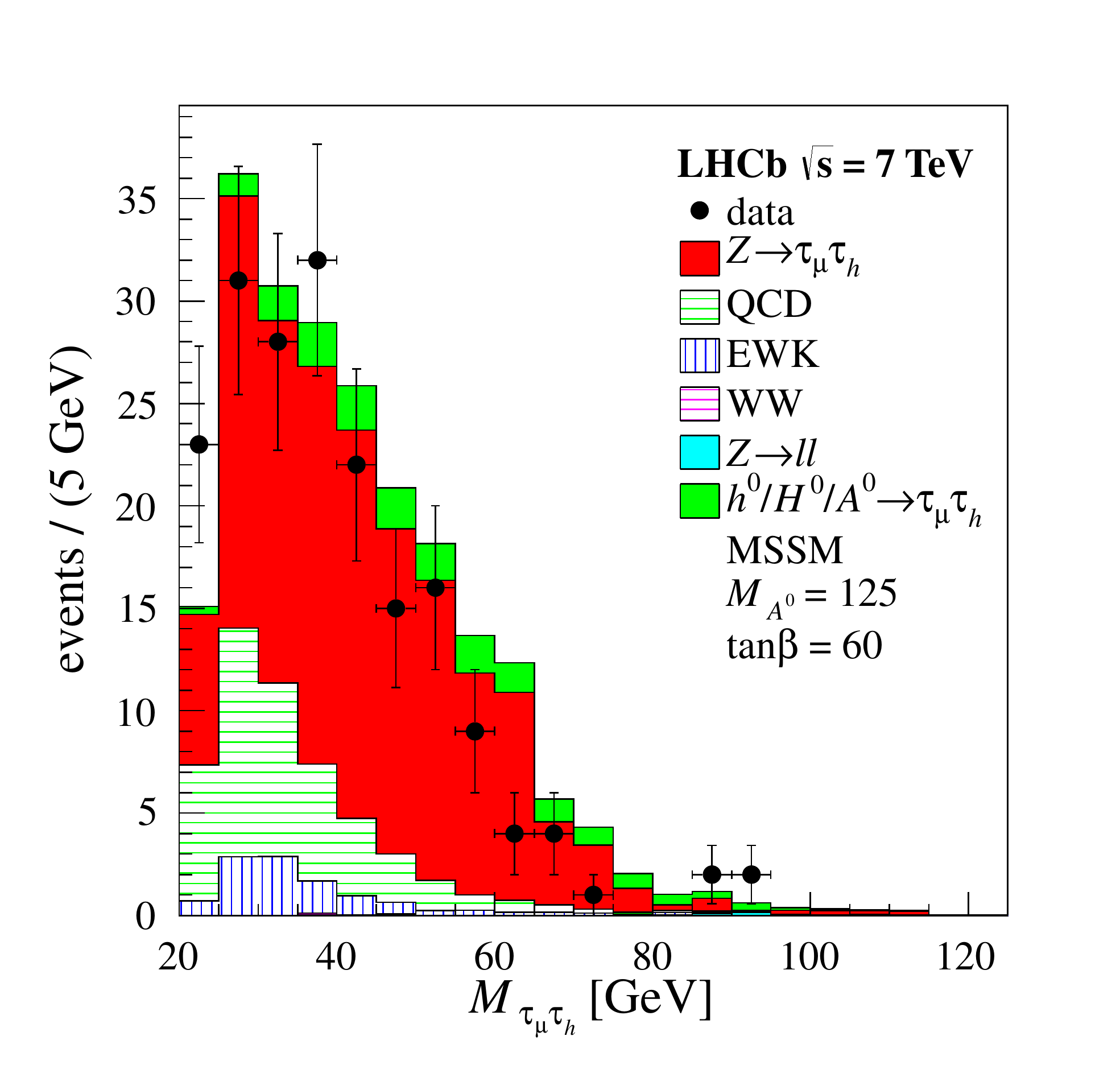}
        \\[\plotsep]
        \subfloat[]{\label{fig:eh}}\hspace{\capwidth} &
        \subfloat[]{\label{fig:all}}\hspace{\capwidth} \\[\capsep]
        \includegraphics[width=\plotwidth]
        {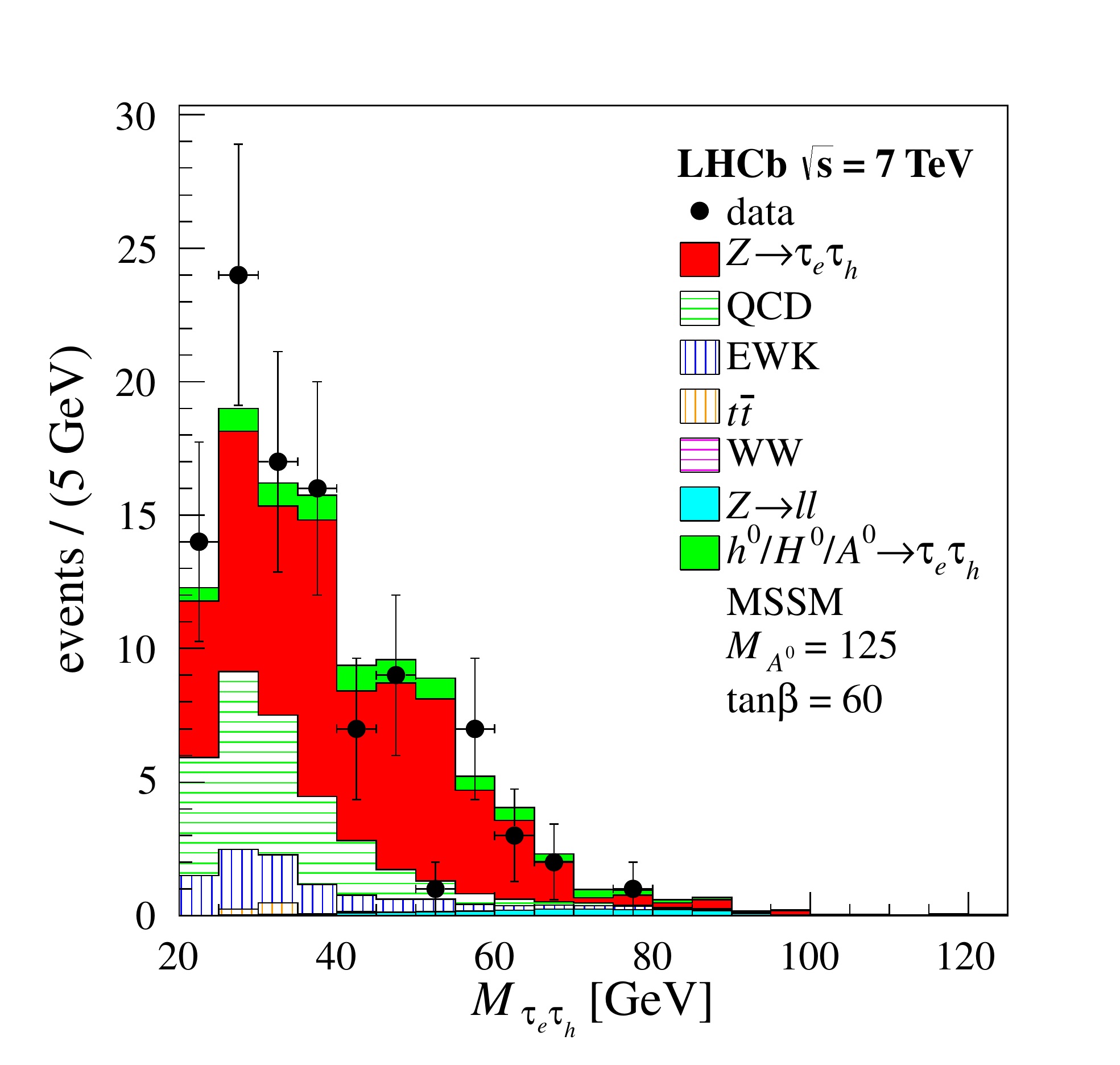} &
        \includegraphics[width=\plotwidth]
        {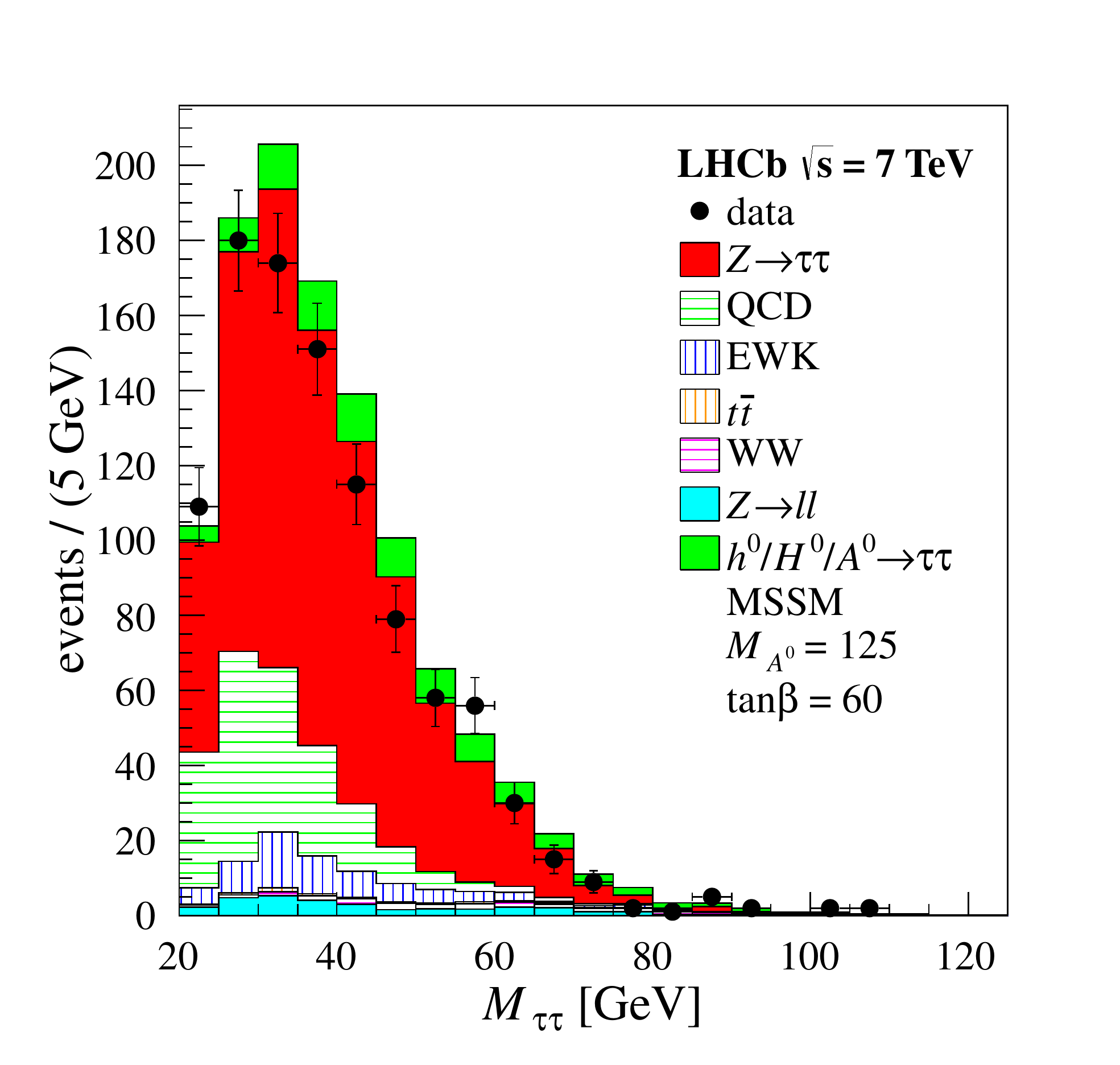} \\
      \end{tabular}}
  \end{center}
  \vspace{-0.95cm}
  \caption{\small Invariant mass distributions for
    \protect\subref{fig:mumu} \mumu, \protect\subref{fig:mue} \mue,
    \protect\subref{fig:emu} \emu, \protect\subref{fig:muh} \muh,
    \protect\subref{fig:eh} \eh, and \protect\subref{fig:all} all
    candidates. The ${\z \to \ditau}$ background (solid red) is
    normalised to the theoretical expectation. The \qcd (horizontal
    green), electroweak (vertical blue), and \z (solid cyan)
    backgrounds are estimated from data. The \ttbar (vertical orange)
    and \ww (horizontal magenta) backgrounds are estimated from
    simulation and generally not visible. The contribution that would
    be expected from an MSSM signal for ${M_\hAz = 125~\gev}$ and
    ${\tan\beta = 60}$ is shown in solid green.\label{fig:data}}
\end{figure}

The invariant mass distribution of the two final state particles for
the selected ${\hz \to \ditau}$ candidates is plotted in
\fig{\ref{fig:data}} for each of the five categories separately and
combined together. No candidates are observed with a mass above
${120~\gev}$. The distributions of \fig{\ref{fig:data}} differ from
those of \rfr{\cite{LhcbZtautau}} as the simulated mass shapes are
calibrated to correct for differences between data and simulation, and
the $\z \to \ditau$ distributions are normalised to theory.

Six background components are considered: ${\z \to \ditau}$; hadronic
processes (QCD); electroweak (EWK), where one $\tau$ decay product
candidate originates from a \w or \z boson and the other comes from
the underlying event; \ttbar; \ww; and ${\z \to \dilep}$ where \dilep
indicates electrons or muons originating from a leptonic \z decay.

All backgrounds, except ${\z \to \ditau}$, have been estimated in
\rfr{\cite{LhcbZtautau}}. The distribution and normalisation of QCD
background events is found from data using same-sign events. The
electroweak invariant mass distribution is taken from simulation and
normalised using data. The small contributions from \ttbar and \ww
production are taken from simulation, while the ${\z \to \dilep}$
invariant mass shape and normalisation are determined from data.

The invariant mass distributions for ${\hz \to \ditau}$ and ${\z \to
  \ditau}$ decays are evaluated from simulation where the mass
resolution has been calibrated using the ${\z \to \dimu}$ invariant
mass peak. Each event is re-weighted by a factor ${(\sigma \times
  \eff{}{})/(\sigma_{\simu} \times \eff{\simu}{})}$, which provides a
negligible correction in comparison to the mass resolution
calibration. The efficiency, \eff{}{}, for triggering, reconstructing
and selecting candidates has been evaluated as a function of momentum
and pseudorapidity using data-driven techniques and is described in
\rfr{\cite{LhcbZtautau}}, while \eff{\simu}{} is the corresponding
efficiency in simulation.  The cross-section for the process in
simulation is represented by $\sigma_{\simu}$, while $\sigma$ is the
theoretical cross-section.  The ${\z \to \ditau}$ sample is normalised
using the cross-section calculated with \dynnlo~\cite{Dynnlo} using
the MSTW08 PDF set~\cite{Mstw}. The ${\hz \to \ditau}$ signal
distribution is found from simulated gluon-fusion events. The signal
samples were generated in mass steps of ${10~\gev}$ from ${90~\gev}$
to $250~\gev$. For both the SM and MSSM Higgs bosons, the
normalisation of the signal uses the theoretical calculations
described below.

The SM cross-sections, using the recommendations of
\rfrs{\cite{HiggsInclusive}} and \cite{HiggsDiff}, are calculated at
${\sqrt{s} = 7~\tev}$ with the program {\sc dfg}~\cite{DFG} in the
complex-pole scheme at next-to-next-to-leading log in QCD
contributions and next-to-leading order (NLO) in electroweak
contributions. The large parameter space in the MSSM necessitates the
use of benchmark scenarios~\cite{Benchmarks}. Only the \mhmax scenario
is considered for comparison with previous results. Both gluon-fusion
and associated ${b\bar{b}}$ production mechanisms are considered; the
former is calculated at NLO in QCD using {\sc Higlu}~\cite{Higlu} with
the top-loop corrected to NNLO using {\sc ggh@nnlo}~\cite{GghAtNnlo},
while the latter is calculated at NNLO in QCD using {\sc
  bbh@nnlo}~\cite{BbhAtNnlo} with the five flavour scheme. For both SM
and MSSM Higgs bosons, the branching fractions are calculated using
\feynhiggs~\cite{FeynHiggs} at the two-loop level.

The expected distributions of background events are displayed in
\fig{\ref{fig:data}} and the estimated numbers of events with their
associated systematic uncertainties, as well as the observed numbers
of candidates from data, are given in \tab{\ref{tab:events}}. The
systematic uncertainty on the ${\z \to \ditau}$ background is
dominated by the statistical uncertainty on the data-driven
determination of the efficiency; the other background uncertainties
are described in \rfr{\cite{LhcbZtautau}}.

\begin{table}
  \caption{\small Estimated number of events for each background component
    and their sum, together with the observed number of candidates and
    the expected number of SM signal events for $M_\hH = 125~\gev$,
    separated by analysis category.\label{tab:events}}
  \begin{center}
    \newlength{\backspace}
    \settowidth{\backspace}{$1$}
    \newcolumntype{E}{>{$}r<{\,\pm$}@{\,}>{$}r<{$}}
    \newcolumntype{F}{>{$}r<{\,\pm$}@{$\,$\hspace{-\backspace}}>{$}r<{$}}
    \begin{tabular}{>{$}l<{$}|E|E|E|F|F}
      & \multicolumn{2}{c|}{\mumu} & \multicolumn{2}{c|}{\mue} &
      \multicolumn{2}{c|}{\emu} & \multicolumn{2}{c|}{\muh} &
      \multicolumn{2}{c}{\eh} \\
      \midrule
      {\z \to \ditau}
      & 79.8&5.6 & 288.2&26.2 & 115.8&12.7 & 146.1&9.7 & \phantom{1}62.1&8.0 \\
      \qcd 
      & 11.7&3.4 & 72.4 &2.2  & 54.0 &3.0  & 41.9 &0.5 & 24.5&0.6 \\
      \ewk
      & 0.0 &3.5 & 40.3 &4.3  & 0.0  &1.3  & 10.8 &0.5 & 9.3 &0.5 \\
      \ttbar
      & <0.1&0.1 & 3.6  &0.4  & 1.0  &0.1  & <0.1 &0.1 & 0.7 &0.4 \\
      \ww 
      & <0.1&0.1 & 13.3 &1.2  & 1.6  &0.2  & 0.2  &0.1 & <0.1&0.1 \\
      {\z \to \dilep}
      & 29.8&7.0 & \multicolumn{2}{c|}{$-$} & \multicolumn{2}{c|}{$-$}
      & 0.4 &0.1 & 2.0&0.2 \\
      \midrule
      {\rm Total}
      & 121.4&10.2 & 417.9&26.7 & 172.4&13.1 & 199.3&\po9.7 & 98.7&\po8.0 \\
      {\rm Observed}
      & \multicolumn{2}{l|}{124} & \multicolumn{2}{l|}{421} 
      & \multicolumn{2}{l|}{155} & \multicolumn{2}{l|}{189}
      & \multicolumn{2}{l}{101} \\
      \midrule
      {\rm SM~Higgs\times100} & 3.9&0.5 & 11.9&1.6 & 3.8&0.5
      & 9.7&1.3 & 4.2&0.6 \\
    \end{tabular}
  \end{center}
\end{table}
\section{Results}

Limits for model independent and MSSM Higgs boson production are
calculated using the method of \rfr{\cite{Cls}} with ${\cls = 95\%}$
and the test statistic of \equ{$14$} from \rfr{\cite{Cowan}}. The test
statistic is defined using the profile extended-likelihood ratio of
the distributions in \fig{\ref{fig:data}}, where the systematic
uncertainties in \tab{\ref{tab:events}} and the uncertainty on the
simulated invariant mass shapes have been incorporated using normally
distributed nuisance parameters. The uncertainty for the invariant
mass shape is determined from the momentum resolution calibration for
simulation, while the primary normalisation uncertainties are from
luminosity determination and the electron reconstruction
efficiency. The distribution of this test statistic is assumed to
follow the result of Wilks~\cite{Wilks}; this assumption has been
validated using a simple likelihood ratio. The expected limits have
been determined using Asimov datasets \cite{Cowan}.

\begin{figure}
  \setlength{\plotwidth}{0.5\textwidth}
  \setlength{\capwidth}{0cm}
  \setlength{\capsep}{-1cm}
  \begin{center}
    \centerline{
      \begin{tabular}{@{\extracolsep{-0.6cm}}cc}
        \includegraphics[width=\plotwidth]
        {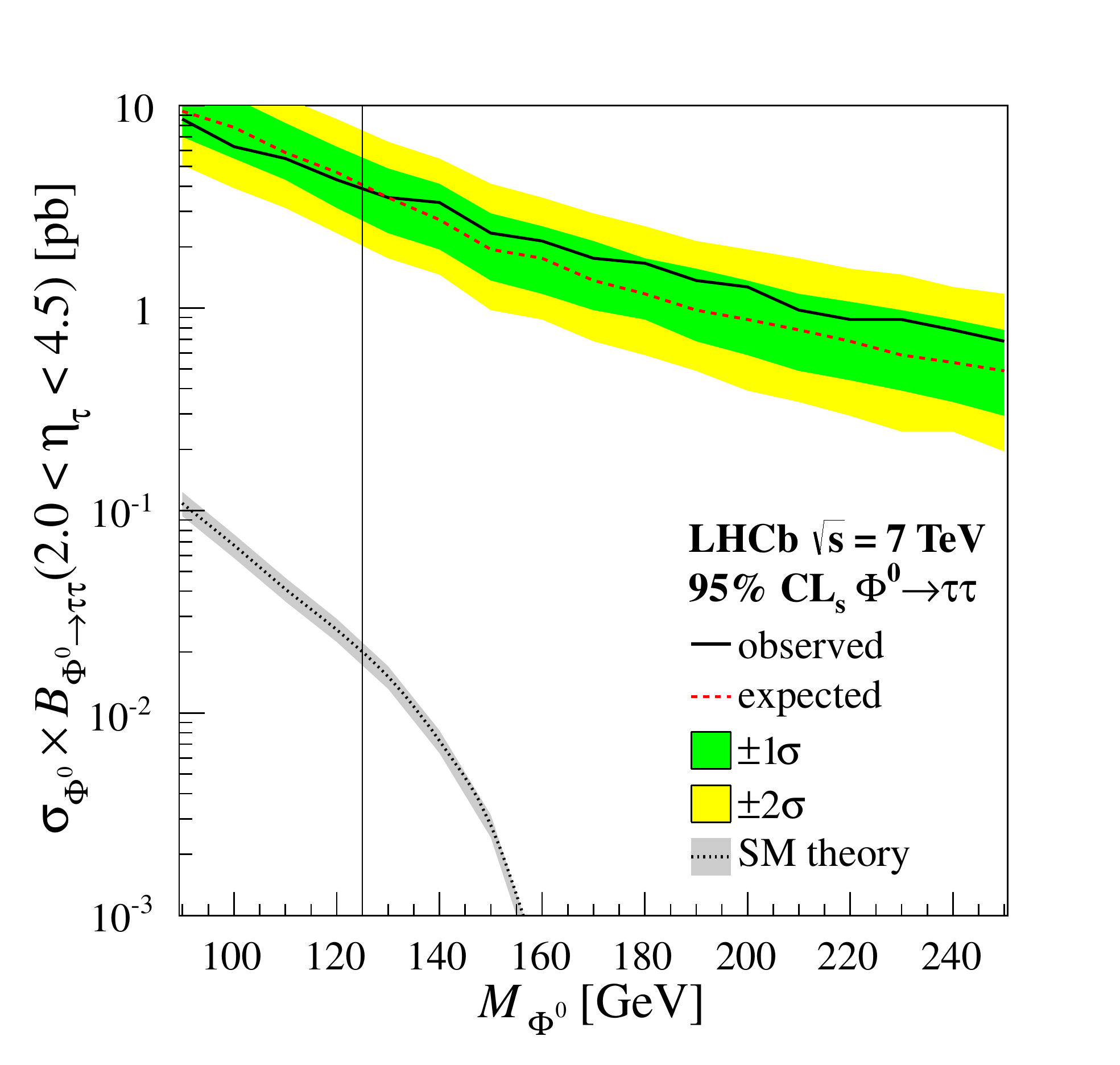} &
        \includegraphics[width=\plotwidth]
        {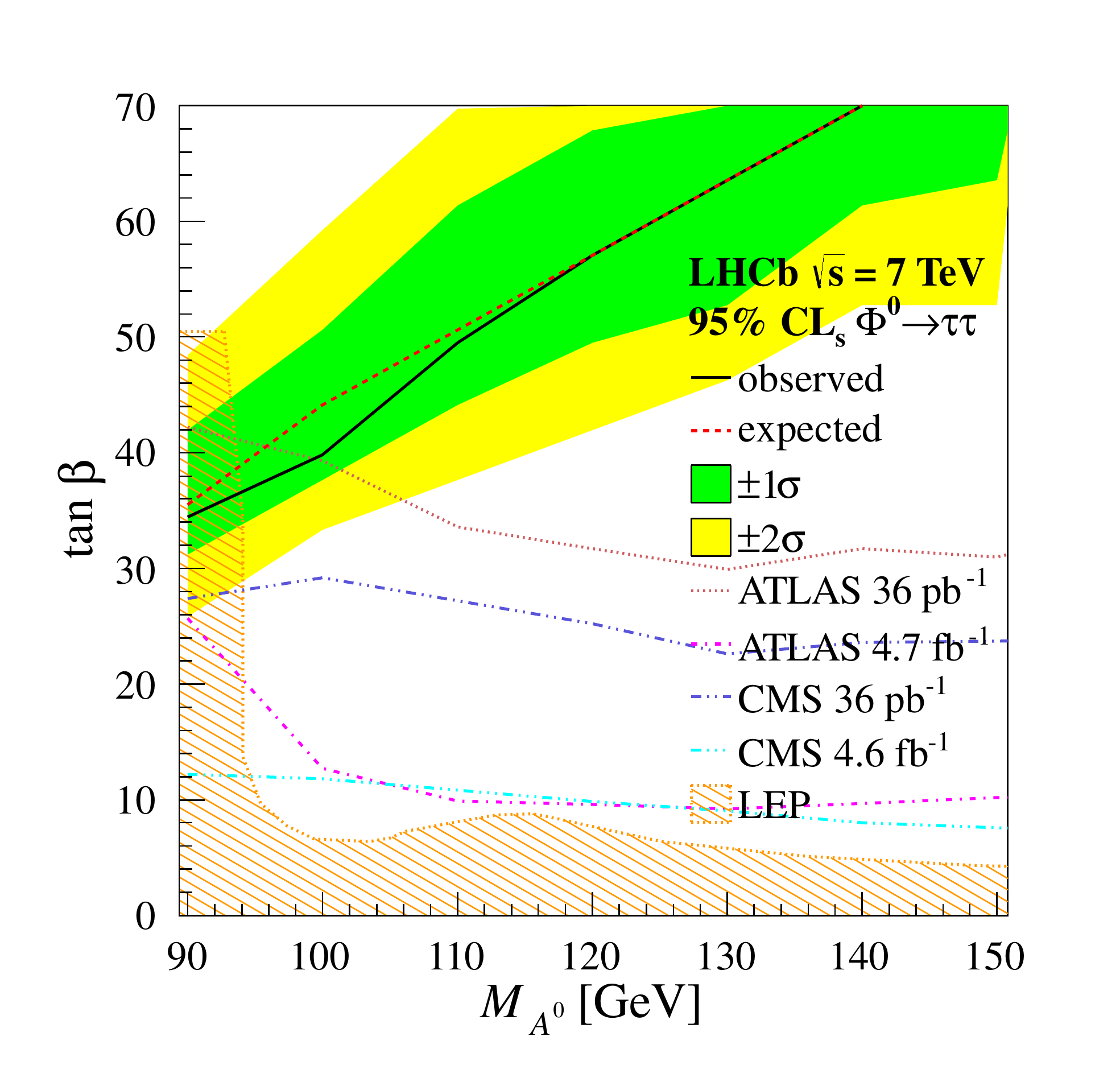} \\
      \end{tabular}}
  \end{center}
  \vspace{-0.95cm}
  \caption{\small Model independent combined limit on cross-section by
    branching fraction for a Higgs boson decaying to two tau leptons
    at ${95\%}$ \cls as a function of $M_\hz$ is given on the left. The
    background only expected limit (dashed red) and ${\pm1\sigma}$
    (green) and ${\pm2\sigma}$ (yellow) bands are compared with the
    observed limit (solid black) and the expected SM theory (dotted
    black) with uncertainty (grey). The combined MSSM ${95\%}$ \cls
    upper limit on ${\tan\beta}$ as a function of $M_\hAz$ is given on
    the right and compared to ATLAS (dotted maroon and dot-dashed
    magenta), CMS (dot-dot-dashed blue and dot-dot-dot-dashed cyan),
    and LEP (hatched orange) results.\label{fig:results}}
\end{figure}

The upper limit on the cross-section times branching fraction of a
model independent Higgs boson decaying to two tau leptons with ${2.0 <
  \eta < 4.5}$ is plotted on the left of \fig{\ref{fig:results}} as a
function of the Higgs boson mass. The upper-limit on ${\tan\beta}$ for
the production of neutral MSSM Higgs bosons, as a function of the
\cp-odd Higgs boson mass, $M_\hAz$, is provided in the right plot of
\fig{\ref{fig:results}}. Previously published exclusion limits from
\atlas~\cite{Atlas2010,Atlas2011}, \cms~\cite{Cms2010,Cms2011}, and
LEP~\cite{Lep} are provided for comparison.

\section{Conclusions}

A model independent search for a Higgs boson decaying to two tau
leptons with pseudorapidities between $2.0$ and $4.5$ gives an upper
bound, at the $95\%$ confidence level, on the cross-section times
branching fraction of $8.6~\pb$ for a Higgs boson mass of ${90~\gev}$
with the bound decreasing smoothly to $0.7~\pb$ for a Higgs boson mass
of ${250~\gev}$.

Limits on a MSSM Higgs bosons have been set in the \mhmax scenario.
Values above ${\tan\beta}$ ranging from $34$ to $70$ are excluded over
the \cp-odd MSSM Higgs boson mass range of $90$ to $140~\gev$. For
${M_{A^0} < 110~\gev}$, these are comparable to the limits obtained by
\atlas and \cms using the $2010$ data sets but are considerably less
stringent than the \atlas and \cms results using $2011$ data. The
forthcoming running of the LHC should allow the boson, observed by
\atlas and \cms, to be seen in the \lhcb detector through a
combination of channels and should provide complementary information
on its properties.

\section*{Acknowledgements}

\noindent We express our gratitude to our colleagues in the CERN
accelerator departments for the excellent performance of the LHC. We
thank the technical and administrative staff at the LHCb
institutes. We acknowledge support from CERN and from the national
agencies: CAPES, CNPq, FAPERJ and FINEP (Brazil); NSFC (China);
CNRS/IN2P3 and Region Auvergne (France); BMBF, DFG, HGF and MPG
(Germany); SFI (Ireland); INFN (Italy); FOM and NWO (The Netherlands);
SCSR (Poland); ANCS/IFA (Romania); MinES, Rosatom, RFBR and NRC
``Kurchatov Institute'' (Russia); MinECo, XuntaGal and GENCAT (Spain);
SNSF and SER (Switzerland); NAS Ukraine (Ukraine); STFC (United
Kingdom); NSF (USA). We also acknowledge the support received from the
ERC under FP7. The Tier1 computing centres are supported by IN2P3
(France), KIT and BMBF (Germany), INFN (Italy), NWO and SURF (The
Netherlands), PIC (Spain), GridPP (United Kingdom). We are thankful
for the computing resources put at our disposal by Yandex LLC
(Russia), as well as to the communities behind the multiple open
source software packages that we depend on.
 
\newboolean{articletitles}
\setboolean{articletitles}{true}
\bibliographystyle{lhcb}
\bibliography{Latex/bibliography}

\ifx\mcitethebibliography\mciteundefinedmacro
\PackageError{LHCb.bst}{mciteplus.sty has not been loaded}
{This bibstyle requires the use of the mciteplus package.}\fi
\providecommand{\href}[2]{#2}
\begin{mcitethebibliography}{10}
\mciteSetBstSublistMode{n}
\mciteSetBstMaxWidthForm{subitem}{\alph{mcitesubitemcount})}
\mciteSetBstSublistLabelBeginEnd{\mcitemaxwidthsubitemform\space}
{\relax}{\relax}

\bibitem{AtlasHiggs}
ATLAS collaboration, \ifthenelse{\boolean{articletitles}}{{\it {Observation of
  a new particle in the search for the Standard Model Higgs boson with the
  ATLAS detector at the LHC}},
  }{}\href{http://dx.doi.org/10.1016/j.physletb.2012.08.020}{Phys.\ Lett.\
  {\bf B716} (2012) 1}, \href{http://arxiv.org/abs/1207.7214}{{\tt
  arXiv:1207.7214}}\relax
\mciteBstWouldAddEndPuncttrue
\mciteSetBstMidEndSepPunct{\mcitedefaultmidpunct}
{\mcitedefaultendpunct}{\mcitedefaultseppunct}\relax
\EndOfBibitem
\bibitem{CmsHiggs}
CMS collaboration, \ifthenelse{\boolean{articletitles}}{{\it {Observation of a
  new boson at a mass of 125 GeV with the CMS experiment at the LHC}},
  }{}\href{http://dx.doi.org/10.1016/j.physletb.2012.08.021}{Phys.\ Lett.\
  {\bf B716} (2012) 30}, \href{http://arxiv.org/abs/1207.7235}{{\tt
  arXiv:1207.7235}}\relax
\mciteBstWouldAddEndPuncttrue
\mciteSetBstMidEndSepPunct{\mcitedefaultmidpunct}
{\mcitedefaultendpunct}{\mcitedefaultseppunct}\relax
\EndOfBibitem
\bibitem{Benchmarks}
M.~Carena, S.~Heinemeyer, C.~Wagner, and G.~Weiglein,
  \ifthenelse{\boolean{articletitles}}{{\it {Suggestions for benchmark
  scenarios for MSSM Higgs boson searches at hadron colliders}},
  }{}\href{http://dx.doi.org/10.1140/epjc/s2002-01084-3}{Eur.\ Phys.\ J.\  {\bf
  C26} (2003) 601}, \href{http://arxiv.org/abs/hep-ph/0202167}{{\tt
  arXiv:hep-ph/0202167}}\relax
\mciteBstWouldAddEndPuncttrue
\mciteSetBstMidEndSepPunct{\mcitedefaultmidpunct}
{\mcitedefaultendpunct}{\mcitedefaultseppunct}\relax
\EndOfBibitem
\bibitem{Detector}
LHCb collaboration, A.~A. Alves~Jr. {\em et~al.},
  \ifthenelse{\boolean{articletitles}}{{\it {The \lhcb detector at the LHC}},
  }{}\href{http://dx.doi.org/10.1088/1748-0221/3/08/S08005}{JINST {\bf 3}
  (2008) S08005}\relax
\mciteBstWouldAddEndPuncttrue
\mciteSetBstMidEndSepPunct{\mcitedefaultmidpunct}
{\mcitedefaultendpunct}{\mcitedefaultseppunct}\relax
\EndOfBibitem
\bibitem{Trigger}
R.~Aaij {\em et~al.}, \ifthenelse{\boolean{articletitles}}{{\it {The \lhcb
  trigger and its performance}}, }{}\href{http://arxiv.org/abs/1211.3055}{{\tt
  arXiv:1211.3055}}, {submitted to JINST}\relax
\mciteBstWouldAddEndPuncttrue
\mciteSetBstMidEndSepPunct{\mcitedefaultmidpunct}
{\mcitedefaultendpunct}{\mcitedefaultseppunct}\relax
\EndOfBibitem
\bibitem{Tune}
I.~Belyaev {\em et~al.}, \ifthenelse{\boolean{articletitles}}{{\it {Handling of
  the generation of primary events in GAUSS, the \lhcb simulation framework}},
  }{}\href{http://dx.doi.org/10.1109/NSSMIC.2010.5873949}{Nuclear Science
  Symposium Conference Record (NSS/MIC) {\bf IEEE} (2010) 1155}\relax
\mciteBstWouldAddEndPuncttrue
\mciteSetBstMidEndSepPunct{\mcitedefaultmidpunct}
{\mcitedefaultendpunct}{\mcitedefaultseppunct}\relax
\EndOfBibitem
\bibitem{Pythia}
T.~Sj\"{o}strand, S.~Mrenna, and P.~Skands,
  \ifthenelse{\boolean{articletitles}}{{\it {PYTHIA 6.4 physics and manual}},
  }{}\href{http://dx.doi.org/10.1088/1126-6708/2006/05/026}{JHEP {\bf 05}
  (2006) 026}, \href{http://arxiv.org/abs/hep-ph/0603175}{{\tt
  arXiv:hep-ph/0603175}}\relax
\mciteBstWouldAddEndPuncttrue
\mciteSetBstMidEndSepPunct{\mcitedefaultmidpunct}
{\mcitedefaultendpunct}{\mcitedefaultseppunct}\relax
\EndOfBibitem
\bibitem{cteq}
J.~Pumplin {\em et~al.}, \ifthenelse{\boolean{articletitles}}{{\it {New
  generation of parton distributions with uncertainties from global QCD
  analysis}}, }{}\href{http://dx.doi.org/10.1088/1126-6708/2002/07/012}{JHEP
  {\bf 07} (2002) 012}, \href{http://arxiv.org/abs/hep-ph/0201195}{{\tt
  arXiv:hep-ph/0201195}}\relax
\mciteBstWouldAddEndPuncttrue
\mciteSetBstMidEndSepPunct{\mcitedefaultmidpunct}
{\mcitedefaultendpunct}{\mcitedefaultseppunct}\relax
\EndOfBibitem
\bibitem{GeantA}
GEANT4 collaboration, J.~Allison {\em et~al.},
  \ifthenelse{\boolean{articletitles}}{{\it {Geant4 developments and
  applications}}, }{}\href{http://dx.doi.org/10.1109/TNS.2006.869826}{IEEE
  Trans.\ Nucl.\ Sci.\  {\bf 53} (2006) 270}\relax
\mciteBstWouldAddEndPuncttrue
\mciteSetBstMidEndSepPunct{\mcitedefaultmidpunct}
{\mcitedefaultendpunct}{\mcitedefaultseppunct}\relax
\EndOfBibitem
\bibitem{GeantB}
GEANT4 collaboration, S.~Agostinelli {\em et~al.},
  \ifthenelse{\boolean{articletitles}}{{\it {GEANT4: a simulation toolkit}},
  }{}\href{http://dx.doi.org/10.1016/S0168-9002(03)01368-8}{Nucl.\ Instrum.\
  Meth.\  {\bf A506} (2003) 250}\relax
\mciteBstWouldAddEndPuncttrue
\mciteSetBstMidEndSepPunct{\mcitedefaultmidpunct}
{\mcitedefaultendpunct}{\mcitedefaultseppunct}\relax
\EndOfBibitem
\bibitem{Gauss}
M.~Clemencic {\em et~al.}, \ifthenelse{\boolean{articletitles}}{{\it {The \lhcb
  simulation application, Gauss: design, evolution and experience}},
  }{}\href{http://dx.doi.org/10.1088/1742-6596/331/3/032023}{{J.\ of Phys.\ :
  Conf.\ Ser.\ } {\bf 331} (2011) 032023}\relax
\mciteBstWouldAddEndPuncttrue
\mciteSetBstMidEndSepPunct{\mcitedefaultmidpunct}
{\mcitedefaultendpunct}{\mcitedefaultseppunct}\relax
\EndOfBibitem
\bibitem{Brunel}
A.~P. Navarro and M.~Frank, \ifthenelse{\boolean{articletitles}}{{\it {Event
  reconstruction in the LHCb online cluster}},
  }{}\href{http://dx.doi.org/10.1088/1742-6596/219/2/022020}{J.\ Phys.\ Conf.\
  Ser.\  {\bf 219} (2010) 022020}\relax
\mciteBstWouldAddEndPuncttrue
\mciteSetBstMidEndSepPunct{\mcitedefaultmidpunct}
{\mcitedefaultendpunct}{\mcitedefaultseppunct}\relax
\EndOfBibitem
\bibitem{LhcbZtautau}
LHCb collaboration, \ifthenelse{\boolean{articletitles}}{{\it {A study of the
  $Z$ production cross-section in $pp$ collisions at $\sqrt{s} = 7$ TeV using
  tau final states}}, }{}\href{http://dx.doi.org/10.1007/JHEP01(2013)111}{JHEP
  {\bf 01} (2013) 111}, \href{http://arxiv.org/abs/1210.6289}{{\tt
  arXiv:1210.6289}}\relax
\mciteBstWouldAddEndPuncttrue
\mciteSetBstMidEndSepPunct{\mcitedefaultmidpunct}
{\mcitedefaultendpunct}{\mcitedefaultseppunct}\relax
\EndOfBibitem
\bibitem{Dynnlo}
S.~Catani and M.~Grazzini, \ifthenelse{\boolean{articletitles}}{{\it {An NNLO
  subtraction formalism in hadron collisions and its application to Higgs boson
  production at the LHC}},
  }{}\href{http://dx.doi.org/10.1103/PhysRevLett.98.222002}{Phys.\ Rev.\ Lett.\
   {\bf 98} (2007) 222002}, \href{http://arxiv.org/abs/hep-ph/0703012}{{\tt
  arXiv:hep-ph/0703012}}\relax
\mciteBstWouldAddEndPuncttrue
\mciteSetBstMidEndSepPunct{\mcitedefaultmidpunct}
{\mcitedefaultendpunct}{\mcitedefaultseppunct}\relax
\EndOfBibitem
\bibitem{Mstw}
A.~D. Martin, W.~J. Stirling, R.~S. Thorne, and G.~Watt,
  \ifthenelse{\boolean{articletitles}}{{\it {Parton distributions for the
  LHC}}, }{}\href{http://dx.doi.org/10.1140/epjc/s10052-009-1072-5}{Eur.\
  Phys.\ J.\  {\bf C63} (2009) 189}, \href{http://arxiv.org/abs/0901.0002}{{\tt
  arXiv:0901.0002}}\relax
\mciteBstWouldAddEndPuncttrue
\mciteSetBstMidEndSepPunct{\mcitedefaultmidpunct}
{\mcitedefaultendpunct}{\mcitedefaultseppunct}\relax
\EndOfBibitem
\bibitem{HiggsInclusive}
LHC Higgs Cross Section Working Group, S.~Dittmaier {\em et~al.},
  \ifthenelse{\boolean{articletitles}}{{\it {Handbook of LHC Higgs cross
  sections: 1. inclusive observables}},
  }{}\href{http://arxiv.org/abs/1101.0593}{{\tt arXiv:1101.0593}}\relax
\mciteBstWouldAddEndPuncttrue
\mciteSetBstMidEndSepPunct{\mcitedefaultmidpunct}
{\mcitedefaultendpunct}{\mcitedefaultseppunct}\relax
\EndOfBibitem
\bibitem{HiggsDiff}
LHC Higgs Cross Section Working Group, S.~Dittmaier {\em et~al.},
  \ifthenelse{\boolean{articletitles}}{{\it {Handbook of LHC Higgs cross
  sections: 2. differential distributions}},
  }{}\href{http://arxiv.org/abs/1201.3084}{{\tt arXiv:1201.3084}}\relax
\mciteBstWouldAddEndPuncttrue
\mciteSetBstMidEndSepPunct{\mcitedefaultmidpunct}
{\mcitedefaultendpunct}{\mcitedefaultseppunct}\relax
\EndOfBibitem
\bibitem{DFG}
D.~de~Florian and M.~Grazzini, \ifthenelse{\boolean{articletitles}}{{\it {Higgs
  production through gluon fusion: updated cross sections at the Tevatron and
  the LHC}}, }{}\href{http://dx.doi.org/10.1016/j.physletb.2009.03.033}{Phys.\
  Lett.\  {\bf B674} (2009) 291}, \href{http://arxiv.org/abs/0901.2427}{{\tt
  arXiv:0901.2427}}\relax
\mciteBstWouldAddEndPuncttrue
\mciteSetBstMidEndSepPunct{\mcitedefaultmidpunct}
{\mcitedefaultendpunct}{\mcitedefaultseppunct}\relax
\EndOfBibitem
\bibitem{Higlu}
M.~Spira, \ifthenelse{\boolean{articletitles}}{{\it {HIGLU: a program for the
  calculation of the total Higgs production cross-section at hadron colliders
  via gluon fusion including QCD corrections}},
  }{}\href{http://arxiv.org/abs/hep-ph/9510347}{{\tt
  arXiv:hep-ph/9510347}}\relax
\mciteBstWouldAddEndPuncttrue
\mciteSetBstMidEndSepPunct{\mcitedefaultmidpunct}
{\mcitedefaultendpunct}{\mcitedefaultseppunct}\relax
\EndOfBibitem
\bibitem{GghAtNnlo}
R.~V. Harlander and W.~B. Kilgore, \ifthenelse{\boolean{articletitles}}{{\it
  {Production of a pseudoscalar Higgs boson at hadron colliders at
  next-to-next-to leading order}},
  }{}\href{http://dx.doi.org/10.1007/JHEP09(2011)088}{JHEP {\bf 10} (2002)
  017}, \href{http://arxiv.org/abs/hep-ph/0208096}{{\tt
  arXiv:hep-ph/0208096}}\relax
\mciteBstWouldAddEndPuncttrue
\mciteSetBstMidEndSepPunct{\mcitedefaultmidpunct}
{\mcitedefaultendpunct}{\mcitedefaultseppunct}\relax
\EndOfBibitem
\bibitem{BbhAtNnlo}
R.~V. Harlander and W.~B. Kilgore, \ifthenelse{\boolean{articletitles}}{{\it
  {Higgs boson production in bottom quark fusion at next-to-next-to leading
  order}}, }{}\href{http://dx.doi.org/10.1103/PhysRevD.68.013001}{Phys.\ Rev.\
  {\bf D68} (2003) 013001}, \href{http://arxiv.org/abs/hep-ph/0304035}{{\tt
  arXiv:hep-ph/0304035}}\relax
\mciteBstWouldAddEndPuncttrue
\mciteSetBstMidEndSepPunct{\mcitedefaultmidpunct}
{\mcitedefaultendpunct}{\mcitedefaultseppunct}\relax
\EndOfBibitem
\bibitem{FeynHiggs}
S.~Heinemeyer, W.~Hollik, and G.~Weiglein,
  \ifthenelse{\boolean{articletitles}}{{\it {FeynHiggs: a program for the
  calculation of the masses of the neutral CP even Higgs bosons in the MSSM}},
  }{}\href{http://dx.doi.org/10.1016/S0010-4655(99)00364-1}{Comput.\ Phys.\
  Commun.\  {\bf 124} (2000) 76},
  \href{http://arxiv.org/abs/hep-ph/9812320}{{\tt arXiv:hep-ph/9812320}}\relax
\mciteBstWouldAddEndPuncttrue
\mciteSetBstMidEndSepPunct{\mcitedefaultmidpunct}
{\mcitedefaultendpunct}{\mcitedefaultseppunct}\relax
\EndOfBibitem
\bibitem{Cls}
A.~L. Read, \ifthenelse{\boolean{articletitles}}{{\it {Presentation of search
  results: the CL(s) technique}},
  }{}\href{http://dx.doi.org/10.1088/0954-3899/28/10/313}{J.\ Phys.\  {\bf G28}
  (2002) 2693}\relax
\mciteBstWouldAddEndPuncttrue
\mciteSetBstMidEndSepPunct{\mcitedefaultmidpunct}
{\mcitedefaultendpunct}{\mcitedefaultseppunct}\relax
\EndOfBibitem
\bibitem{Cowan}
G.~Cowan, K.~Cranmer, E.~Gross, and O.~Vitells,
  \ifthenelse{\boolean{articletitles}}{{\it {Asymptotic formulae for
  likelihood-based tests of new physics}},
  }{}\href{http://dx.doi.org/10.1140/epjc/s10052-011-1554-0}{Eur.\ Phys.\ J.\
  {\bf C71} (2011) 1554}, \href{http://arxiv.org/abs/1007.1727}{{\tt
  arXiv:1007.1727}}\relax
\mciteBstWouldAddEndPuncttrue
\mciteSetBstMidEndSepPunct{\mcitedefaultmidpunct}
{\mcitedefaultendpunct}{\mcitedefaultseppunct}\relax
\EndOfBibitem
\bibitem{Wilks}
S.~S. Wilks, \ifthenelse{\boolean{articletitles}}{{\it {The large-sample
  distribution of the likelihood ratio for testing composite hypotheses}},
  }{}\href{http://www.\ jstor.\ org/stable/2957648}{The Annals of Mathematical
  Statistics} {\bf \href{http://www.jstor.org/stable/2957648}{9
  $\mathrm{No.}~1$}} (\href{http://www.jstor.org/stable/2957648}{1938})
  \href{http://www.jstor.org/stable/2957648}{60}\relax
\mciteBstWouldAddEndPuncttrue
\mciteSetBstMidEndSepPunct{\mcitedefaultmidpunct}
{\mcitedefaultendpunct}{\mcitedefaultseppunct}\relax
\EndOfBibitem
\bibitem{Atlas2010}
ATLAS collaboration, \ifthenelse{\boolean{articletitles}}{{\it {Search for
  neutral MSSM Higgs bosons decaying to {$\tau^+ \tau^-$} pairs in
  proton-proton collisions at {$\sqrt{s}=7$} TeV with the ATLAS detector}},
  }{}\href{http://dx.doi.org/10.1016/j.physletb.2011.10.001}{Phys.\ Lett.\
  {\bf B705} (2011) 174}, \href{http://arxiv.org/abs/1107.5003}{{\tt
  arXiv:1107.5003}}\relax
\mciteBstWouldAddEndPuncttrue
\mciteSetBstMidEndSepPunct{\mcitedefaultmidpunct}
{\mcitedefaultendpunct}{\mcitedefaultseppunct}\relax
\EndOfBibitem
\bibitem{Atlas2011}
ATLAS collaboration, \ifthenelse{\boolean{articletitles}}{{\it {Search for the
  neutral Higgs bosons of the Minimal Supersymmetric Standard Model in $pp$
  collisions at $\sqrt{s}=7$ TeV with the ATLAS detector}},
  }{}\href{http://dx.doi.org/10.1007/JHEP02(2013)095}{JHEP {\bf 02} (2013)
  095}, \href{http://arxiv.org/abs/1211.6956}{{\tt arXiv:1211.6956}}\relax
\mciteBstWouldAddEndPuncttrue
\mciteSetBstMidEndSepPunct{\mcitedefaultmidpunct}
{\mcitedefaultendpunct}{\mcitedefaultseppunct}\relax
\EndOfBibitem
\bibitem{Cms2010}
CMS collaboration, \ifthenelse{\boolean{articletitles}}{{\it {Search for
  neutral MSSM Higgs bosons decaying to tau pairs in $pp$ collisions at
  $\sqrt{s}=7$ TeV}},
  }{}\href{http://dx.doi.org/10.1103/PhysRevLett.106.231801}{Phys.\ Rev.\
  Lett.\  {\bf 106} (2011) 231801}, \href{http://arxiv.org/abs/1104.1619}{{\tt
  arXiv:1104.1619}}\relax
\mciteBstWouldAddEndPuncttrue
\mciteSetBstMidEndSepPunct{\mcitedefaultmidpunct}
{\mcitedefaultendpunct}{\mcitedefaultseppunct}\relax
\EndOfBibitem
\bibitem{Cms2011}
CMS collaboration, \ifthenelse{\boolean{articletitles}}{{\it {Search for
  neutral Higgs bosons decaying to $\tau$ pairs in $pp$ collisions at
  $\sqrt{s}=7$ TeV}},
  }{}\href{http://dx.doi.org/10.1016/j.physletb.2012.05.028}{Phys.\ Lett.\
  {\bf B713} (2012) 68}, \href{http://arxiv.org/abs/1202.4083}{{\tt
  arXiv:1202.4083}}\relax
\mciteBstWouldAddEndPuncttrue
\mciteSetBstMidEndSepPunct{\mcitedefaultmidpunct}
{\mcitedefaultendpunct}{\mcitedefaultseppunct}\relax
\EndOfBibitem
\bibitem{Lep}
ALEPH collaboration, DELPHI collaboration, L3 collaboration, OPAL
  collaboration, LEP Working Group for Higgs Boson Searches, S.~Schael {\em
  et~al.}, \ifthenelse{\boolean{articletitles}}{{\it {Search for neutral MSSM
  Higgs bosons at LEP}},
  }{}\href{http://dx.doi.org/10.1140/epjc/s2006-02569-7}{Eur.\ Phys.\ J.\  {\bf
  C47} (2006) 547}, \href{http://arxiv.org/abs/hep-ex/0602042}{{\tt
  arXiv:hep-ex/0602042}}\relax
\mciteBstWouldAddEndPuncttrue
\mciteSetBstMidEndSepPunct{\mcitedefaultmidpunct}
{\mcitedefaultendpunct}{\mcitedefaultseppunct}\relax
\EndOfBibitem
\end{mcitethebibliography}

\end{document}